\documentclass[aps,prd,10pt,nofootinbib,twocolumn]{revtex4-1}
\usepackage{amsmath,amssymb,amsfonts,dsfont,mathrsfs,amsthm,mathtools}
\usepackage{hyperref}
\usepackage[english]{babel}
\hypersetup{linktocpage,colorlinks=true,urlcolor=blue,linkcolor=blue,citecolor=blue}
\DeclareMathOperator\arcsinh{arcsinh}
\usepackage[pdftex]{graphicx}
\usepackage{pgf,tikz,wasysym}
\usetikzlibrary{arrows}

\begin{document}

\title{Energy nonconservation and relativistic trajectories: Unimodular gravity and beyond}

\author{Y. Bonder}
\email{bonder@nucleares.unam.mx}
\author{J.E. Herrera}
\author{A.M. Rubiol}
\affiliation{Instituto de Ciencias Nucleares, Universidad Nacional Aut\'onoma 
de M\'exico\\ Apartado Postal 70-543, Cd.~Mx., 04510, M\'exico}

\begin{abstract}
Energy conservation has the status of a fundamental physical principle. However, measurements in quantum mechanics do not comply with energy conservation. Therefore, it is expected that a more fundamental theory of gravity ---one that is less incompatible with quantum mechanics--- should admit energy nonconservations. This paper begins by identifying the conditions for a theory to have an energy-momentum tensor that is not conserved. Then, the trajectory equation for pointlike particles that lose energy is derived, showing that energy nonconservation produces a particular acceleration. As an example, the unimodular theory of gravity is studied. Interestingly, in spherical symmetry, given that there is a generalized Birkhoff theorem and that the energy-momentum tensor divergence is a closed form, the trajectories of test particles that lose energy can be found using well known methods. Finally, limits on the energy nonconservation parameters are set using Solar system observations.
\end{abstract}

\maketitle

\section{Introduction}

Energy conservation is regarded as one of the most fundamental pillars in physics \cite{Lindsay}. However, energy can fail to be conserved during a quantum-mechanical measurement \cite{MAUDLIN202067}. An evident example is given by a wave packet that is wide in energy. When an energy measurement is performed, the state collapses into an energy eigenstate, and the hamiltonian expectation value will, most likely, change. Usually, it is argued that energy conservation is restored when considering the interaction with the (classical) measuring apparatus. However, using entangled states, it is easy to devise examples where this cannot occur, since the energy exchange with the measuring apparatus can be arbitrarily small when compared to the change in the hamiltonian expectation value \cite{MAUDLIN202067}. What is more, in several quantum theories that try to solve the measurement problem, like most Dynamical Reduction Models \cite{GRW,Bassi}, energy is explicitly not conserved, and similar issues show up in semiclassical gravity \cite{DiezSudarsky}.

Nowadays, General relativity (GR) is the paradigmatic theory for gravitation. In GR, there is no generic notion of gravitational energy, and thus, no energy conservation law can be proposed \cite{Bondi} (see Refs.~\onlinecite{Katz_2005,Katz_2006} for examples of situations where gravitational energy can be defined). Yet, the energy of matter\footnote{Matter refers to all nongravitational fields.} is conserved in the sense that the energy-momentum tensor has vanishing divergence. In turn, this is a consequence diffeomorphism invariance and it played a key role in the construction of GR \cite{EinsteinAndEnergyConservation}.

One of the main challenges of modern physics is to consistently couple gravity with quantum-mechanical matter \cite{oriti_2009}. This, together with the prediction of spacetime singularities \cite{Singularities}, and the necessity to add dark matter \cite{DarkMatter} and dark energy \cite{RevModPhys.75.559}, strongly suggests that GR may be replaced by a more fundamental theory. This has motivated the community to propose alternative theories of gravity, some of which remain in the framework of pseudoriemannian geometry.

The most popular alternative theories of gravity lie within the $ \mathfrak{f}(R)$ framework \cite{fofR}, which gives rise to conventional matter energy conservation. However, there are other mechanisms to construct alternative theories of gravity that lead to energy nonconservation. In fact, one could argue that theories with unconventional conservation laws should be ``less incompatible'' with quantum mechanics. As such, it is relevant to verify if energy nonconservation can be incorporated into the relativistic framework: this is the main goal of the paper.

For concreteness, attention is set on the propagation of pointlike particles that do not comply with energy conservation. Of course, pointlike particles play a key role in the construction of any geometrical theory \cite{Operational1,Operational2}, and, in GR, it is well understood how these particles lose energy through gravitational radiation \cite{Gralla_2008,Gralla_2011,Barack_2009}. In this paper, the energy-loss mechanism is assumed to be general and should be thought to arise from more fundamental physics. Also, notice that gravitational backreaction is not \textit{a priori} neglected, even though it is not considered in the examples in the last part of the paper.

Throughout the text, the notation and conventions of Ref.~\onlinecite{wald1984general} are followed. Use is made of abstract indexes, which are denoted with lowercase Latin characters from the beginning of the alphabet: $a,b,c,\ldots$. Component indexes are represented by Greek characters. As is customary, a pair of repeated indexes indicates the corresponding contraction. Moreover, the spacetime metric is denoted by $g_{ab}$, and $g$, without indexes, stands for the determinant of $g_{\mu\nu}$. When convenient, indexes are raised (lowered) with the inverse metric $g^{ab}$ ($g_{ab}$). The metric-compatible and torsion-free derivative operator is $\nabla_a$ and ${R_{abc}}^d$ is the Riemann tensor associated with it. The Ricci tensor is $R_{ab}\equiv {R_{acb}}^c$ and $R\equiv g^{ab}R_{ab}$ is the curvature scalar. Finally, pairs of indexes in between parenthesis represent its symmetrical part weighted by $1/2$. Units where $G=1=c$ are used, however, $c$ is reintroduced when it helps to make certain approximations.

A description of the structure of the paper may be useful: in the next section, the conditions for a theory to have a divergence free energy-momentum tensor are deduced. In Sec.~\ref{SecPapapetrou}, the equation for the trajectory of a pointlike particle that loses energy is derived. To make concrete calculations, the assumptions of staticity and spherical symmetric are considered in Sec.~\ref{StaticSpherical}. Sec.~\ref{SecUnimodular} deals with the application of the formalism to the unimodular theory of gravity, and, in Sec.~\ref{Empirical}, some empirical bounds are set using Solar system data. The concluding remarks are presented in Sec.~\ref{Conclusions}.

\section{When is the energy-momentum tensor divergence free?}

This section is devoted to the conditions that lead to a vanishing divergence of the energy-momentum tensor. It is assumed that gravity is geometrical, that spacetime is four dimensional, and that matter is described by conventional tensor fields\footnote{An example of unconventional fields are bitensorial fields \cite{YuriJohas}.}, which are standard assumptions in most alternative theories of gravity. The additional hypotheses under consideration are:
\begin{enumerate}
\item Gravity is completely described by the metric.
\item Minimal coupling: the total action can be naturally separated as
\begin{equation}
S[g,\psi]= \frac{1}{2\kappa}S_G [g]+ S_M [g,\psi],
\end{equation}
where $\kappa$ is the gravitational coupling constant. Also, $S_G$ and $S_M$ are respectively known as the gravitational and matter actions.
\item The theory is invariant under general diffeomorphisms. In particular, $S_M$ is invariant under such transformations.
\end{enumerate}
Hypothesis 1 implies that the relevant derivative operator is $\nabla_a$, which, being metric compatible and torsion free, is determined by $g_{ab}$ \cite[chapter 3.1]{wald1984general}. Moreover, hypothesis 3 is closely related to the assumption that all the fields in the action are dynamical \cite{Cristobal1} and it may be regarded as an application of the principle of general covariance \cite{wald1984general}.

Under these hypotheses, invariance of the matter action under a generic infinitesimal diffeomorphism produces
\begin{equation} \label{actionvariation}
0=\delta S_M [g,\psi] = \int d^4 x \left( \frac{\delta \mathcal{L}_M }{\delta g^{ab}}\delta g^{ab} + \frac{\delta\mathcal{L}_M  }{\delta \psi}\delta \psi\right),
\end{equation}
where $\mathcal{L}_M$ is the matter lagrangian density. It is clear that, on shell, $\delta \mathcal{L}_M /\delta \psi =0$. In addition, the energy-momentum tensor is defined as
\begin{equation}
T_{ab}\equiv -\frac{2}{\sqrt{-g}}\frac{\delta \mathcal{L}_M }{\delta g^{ab}}=T_{(ab)}.\label{Tab}
\end{equation}
Since the action variation is taken with respect to an infinitesimal diffeomorphism along an arbitrary vector field $\xi^a$, the inverse metric variation is the corresponding Lie derivative, namely, $\delta g^{ab}= -2 \nabla^{(a} \xi^{b)}$. Inserting these results into Eq.~\eqref{actionvariation} leads to
\begin{equation}
0=- \int d^4 x \sqrt{-g}\xi^{b} \nabla^{a} T_{ab} ,\label{actionvariation1}
\end{equation}
where an integration by parts is performed and the boundary term is ignored; these terms are disregarded throughout the manuscript. Since Eq.~\eqref{actionvariation1} is valid for all $\xi^a$, it follows that $\nabla^{a} T_{ab} =0$. Therefore, any theory that satisfies the three listed hypotheses gives rise to a divergence free energy-momentum tensor. In particular, Lovelock's theorem \cite{Lovelock} implies that GR with a cosmological constant is the only theory that satisfies that above listed conditions and that leads to second order equations of motion. On the other hand, $ \mathfrak{f}(R)$ theories, where $S_G= \int d^4 x \sqrt{-g}\  \mathfrak{f}(R)$ for a given function $ \mathfrak{f}$, satisfy the listed hypotheses, and therefore, the corresponding energy-momentum tensor is conserved.

This work is devoted to theories where at least one of the listed hypotheses is not met. In such cases, a generalization of the previous argument can be used to find the modified conservation law; these laws are referred to as ``energy nonconservation'' conditions. A particular example for when hypothesis 1 is negated is to consider an independent connection $C_{ab}^c$. The matter action variation, in this case, takes the form
\begin{eqnarray}
0 &=&\delta S_M [g,C,\psi] \nonumber\\
&=& \int d^4 x \left( \frac{\delta \mathcal{L}_M }{\delta g^{ab}}\delta g^{ab}+\frac{\delta \mathcal{L}_M }{\delta C_{ab}^c}\delta C_{ab}^c + \frac{\delta \mathcal{L}_M }{\delta \psi}\delta \psi\right)\nonumber\\
&=&- \int d^4 x\sqrt{-g}\xi^{d} \left[ \nabla^{a} T_{ad} + \Sigma^{ab}_{c} \nabla_d C^c_{ab} \right.\nonumber\\
&& \left. + \nabla_c(\Sigma^{ab}_{d} C_{ab}^c) - \nabla_a(\Sigma^{ab}_{c} C_{db}^c) - \nabla_b(\Sigma^{ab}_{c} C_{ad}^c) \right],\nonumber\\
&&\label{actionvariation2}
\end{eqnarray}
where $\Sigma^{ab}_c\equiv -(1/\sqrt{-g})(\delta \mathcal{L}_M/\delta C_{ab}^c)$, and $\delta C_{ab}^c$ is given by the Lie derivative of $C_{ab}^c$, which is expressed in terms of $\nabla_a$. Clearly, for Eq.~\eqref{actionvariation2} to be valid for all $\xi^a$, it is necessary that what is inside the brackets vanishes, leading to the corresponding energy nonconservation law. Notice that the key condition for energy nonconservation, in this case, is that the matter action depends on additional gravitational degrees of freedom. Other theories with additional gravitational degrees of freedom include the DHOST theories \cite{DHOST}, among others.

When hypothesis 2 is invalid, there is no clear definition of $T_{ab}$, and hence, no reason for it to have a vanishing divergence. Still, there are theories, like that of Ref.~\onlinecite{PhysRevD.78.064036}, where there is a natural energy-momentum tensor that does not have vanishing divergence. Finally, when invariance under diffeomorphisms is explicitly broken, $\xi^a$ is not general, and it is impossible to conclude, from Eq.~\eqref{actionvariation1}, that $\nabla_a T^{ab}=0$. Restricted diffeomorphisms appear in theories with nondynamical fields, such as the unimodular theory of gravity or in the parametrization of Lorentz violation known as the Standard Model Extension \cite{SME1,SME2,Kostelecky2004} (for concrete examples see Refs.~\onlinecite{PhysRevD.88.105011,Cristobal2,Explicit,PhysRevD.101.064056}). In fact, to produce a conventional energy conservation law, many studies in the context of the  Standard Model Extension assume that Lorentz and diffeomorphism invariance are spontaneously broken \cite{Kostelecky2004,KosteleckyBlumm1,KosteleckyBlumm2,Bonder2015,Gabriel,PhysRevD.103.104016}.

\section{Pointlike particles}\label{SecPapapetrou}

In this section, the equation for the trajectory of a pointlike particle that loses energy is derived following Papapetrou's method \cite{papapetrou}. It is assumed that the particle's energy loss is given by $\tilde{j}^{b}\equiv \nabla_{a}\tilde{t}^{ab}$, where $\tilde{t}^{ab}\equiv\sqrt{-g}t^{ab}$ is the tensor density associated with the particle's energy-momentum tensor, $t^{ab}=t^{ba}$. Note that, in this section, the spacetime metric is considered to be known.

\begin{figure}
\begin{center}
 \includegraphics[width=0.5\textwidth]{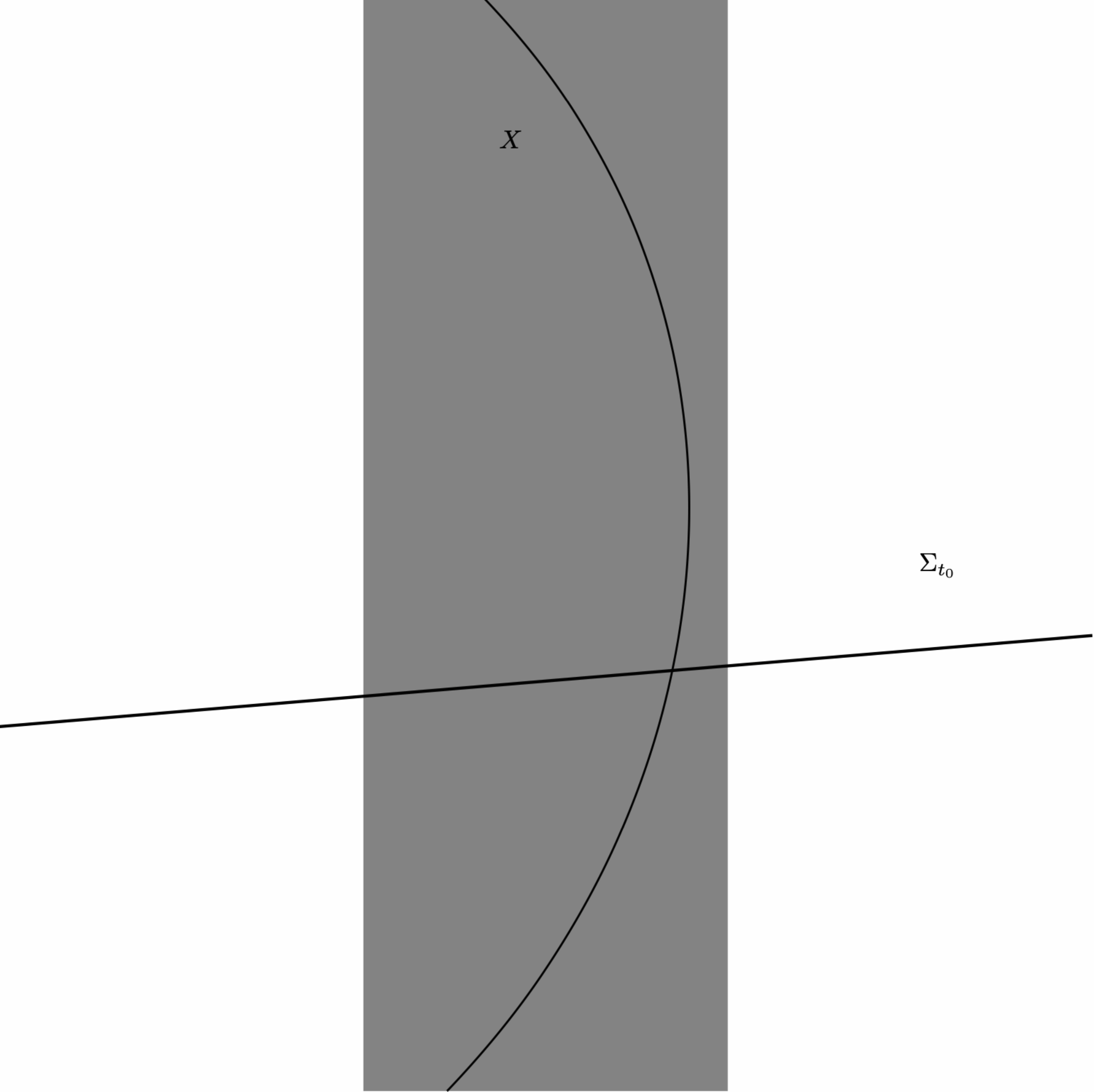}
 \caption{\label{FiguraPapapetrou}Papapetrou's method ($t$-$x^1$ plane). The gray region represents the world tube of the particle that is given by the support of its energy-momentum tensor. The arbitrary curve $X$ and the $t=t_0$ hypersurface, $\Sigma_{t_0}$, are depicted.}
 \end{center}
 \end{figure}

The method begins by assuming that the particle is much smaller than the characteristic radius of the gravitational field, where the ``size'' of the particle is extracted from the support of $t^{ab}$. Eventually, the limit where the size of the particle goes to zero is considered, which eliminates the dependence on the foliation and the spatial coordinates. As the particle evolves, it traces a world tube. Within this world tube, an arbitrary smooth temporal curve $X$ is chosen. Fermi normal coordinates associated with $X$ are used\footnote{These Fermi normal coordinates are defined as follows: The curve $X$ is affinely parametrized by $t$, and an arbitrary orthonormal basis, whose timelike vector is parallel to the curve tangent, $u^a$, is specified at a certain point on $X$. Then, this basis is transported along $X$ in such a way that the basis is kept orthonormal and its timelike vector always coincides with $u^a$ \cite{poisson_2004}. Finally, to give coordinates to a point $q$, it is necessary to find the point $p \in X$ and the vector $v^a$ at $p$ such that $g_{ab} u^a v^b = 0 $ and the geodesic that emanates from $p$ with tangent $v^a$ ``lands'' on $q$ after an affine distance $1$. The coordinates of $q$ are $x^0=t$ and the three spatial coordinates, $x^1$, $x^2$, and $x^3$, are the nonzero components of $v^a$ in the transported basis. Notably, there always exists a neighborhood of $X$ where this coordinates are well defined \cite{poisson_2004}, and here it is assumed that such a region contains the particle's world tube.}. Inside the support of $t^{ab}$, these coordinates generate a foliation by constant $t$ hypersurfaces, $\Sigma_{t}$ (see Fig.~\ref{FiguraPapapetrou}). In what remains of this section, these coordinates are utilized, as it is emphasized by the use of Greek indexes.

From its definition, it is possible to see that
\begin{equation}
\tilde{j}^{\alpha}=\partial_{\mu}\tilde{t}^{ \alpha\mu}+ \Gamma^{\alpha}_{\mu \nu}\tilde{t}^{\mu \nu}.\label{j}
\end{equation}
Multiplying by $x^\beta$, taking the symmetric part, and adding $\tilde{t}^{\alpha \beta}$, yields
\begin{equation}
\tilde{t}^{\alpha \beta}+ x^{(\alpha} \tilde{j}^{\beta)}=\tilde{t}^{\alpha \beta} + x^{(\alpha} \partial_{\mu}\tilde{t}^{ \beta)\mu}+ x^{(\alpha}\Gamma^{\beta)}_{\mu \nu}\tilde{t}^{\mu \nu}.
\end{equation}
Introducing a Kronecker delta as $\partial_\mu x^{\alpha}$, the last equation takes the form
\begin{equation}
\partial_\mu (x^{(\alpha}\tilde{t}^{ \beta)\mu} )=\tilde{t}^{\alpha \beta}+ x^{(\alpha} \tilde{j}^{\beta)}- x^{(\alpha}\Gamma^{\beta)}_{\mu \nu}\tilde{t}^{\mu \nu}.\label{j1}
\end{equation}
Integrating Eqs.~\eqref{j} and \eqref{j1} in $\Sigma_t$ yields
\begin{eqnarray}
\frac{d}{dt}\int d^3x\ \tilde{t}^{t \alpha}&=&\int d^3x\left[ \tilde{j}^{\alpha}- \tilde{t}^{\mu \nu}\Gamma^{\alpha}_{\mu \nu}\right],\label{Pap1}
\\
\frac{d}{dt}\int d^3x\ x^{(\alpha}\tilde{t}^{ \beta)t} &=&\nonumber\\
&\int d^3x &\left[\tilde{t}^{\alpha \beta}+ x^{(\alpha} \tilde{j}^{\beta)}- x^{(\alpha}\Gamma^{\beta)}_{\mu \nu}\tilde{t}^{\mu \nu}\right],\nonumber\\\label{Pap2}
\end{eqnarray}
where the integrals of the spatial divergences are zero by virtue of the divergence theorem and the fact that the integrands have compact support, and the index $t$ refers to the component along the $t$ coordinate. Recall that the integrands are densities, and thus, the integrals are well defined.

Importantly, in the point particle approximation (also known as the monopole approximation), which is the relevant regime for this work, it is enough to work to zeroth order in $\delta x^{\alpha}\equiv x^{\alpha}- X^{\alpha}$. Eqs.~\eqref{Pap1}-\eqref{Pap2}, to zeroth order in $\delta x^{\alpha}$, become
\begin{equation}
\frac{d}{dt}\int d^3x\ \tilde{t}^{t \alpha}=\int d^3x\ \tilde{j}^{\alpha}- \Gamma^{\alpha}_{\mu \nu}(X)\int d^3x\ \tilde{t}^{\mu \nu},\label{Pap1-1}
\end{equation}
and
\begin{eqnarray}
&& \frac{dX^{(\alpha}}{dt} \int d^3x\ \tilde{t}^{ \beta)t}+X^{(\alpha}\frac{d}{dt}\int d^3x\ \tilde{t}^{ \beta)t}\nonumber \\ 
&=& \int d^3x \ \tilde{t}^{\alpha \beta}+ X^{(\alpha} \int d^3x \ \tilde{j}^{\beta)}\nonumber\\
&&- X^{(\alpha}\Gamma^{\beta)}_{\mu \nu}(X) \int d^3x \ \tilde{t}^{\mu \nu}.
\end{eqnarray}
These equations can be combined as
\begin{equation}
\int d^3x \ \tilde{t}^{\alpha \beta}=\frac{dX^{(\alpha}}{dt} \int d^3x\ \tilde{t}^{ \beta)t} .\label{Pap2-2}
\end{equation}

Recall that $t$ is an affine parameter along $X$, and thus, $u^\alpha \equiv d X^\alpha/d t $ is such that $u^\alpha u_\alpha=-1$ and $u^t=1$. What is more, it is convenient to define $M^{\alpha\beta}\equiv  \int d^3x \ \tilde{t}^{\alpha \beta} $. Then, Eq.~\eqref{Pap2-2} takes the form
\begin{equation}
M^{\alpha \beta}=u^{(\alpha}M^{ \beta)t} ,\label{Pap3}
\end{equation}
and its $\alpha t$ component satisfies
\begin{eqnarray}
M^{\alpha t}=u^{\alpha}M^{ tt}.\label{Mresult}
\end{eqnarray}
By means of Eq.~\eqref{Mresult}, Eq.~\eqref{Pap3} can be written as
\begin{equation}
M^{\alpha \beta}= m u^{\alpha}u^{\beta} ,\label{Pap4}
\end{equation}
with $m\equiv M^{tt}$, which is taken as positive invoking the weak energy condition \cite{wald1984general}.
 
With these results, Eq.~\eqref{Pap1-1} becomes
\begin{equation}
\frac{u^{\alpha}}{m}\frac{d m}{dt}+u^\beta \nabla_\beta u^\alpha =J^\alpha,\label{Pap3-1}
\end{equation}
where $J^\alpha\equiv (1/m)\int d^3x\ \tilde{j}^{\alpha}$. Multiplying by $u_\alpha$ yields
\begin{equation}
\frac{-1}{m}\frac{dm}{dt} =J^\alpha u_\alpha,
\end{equation}
where use is made of the fact that $u_\alpha u^\beta \nabla_\beta u^\alpha =  (1/2)u^\beta \nabla_\beta(-1)=0$. With all this, Eq.~\eqref{Pap3-1} can be written as
\begin{equation}
u^\beta \nabla_\beta u^\alpha = J_\perp^{\alpha},\label{PapFinal}
\end{equation}
with $J_\perp^{\alpha}\equiv J^{\beta}(\delta^\alpha_\beta+u^\alpha u_\beta)$ being components of the projection of $J^{a}$ orthogonal to $u^a$. Notice that a projection is expected by the fact that $t$ is an affine parameter. In addition, when $J_\perp^{\alpha}=0$, the geodesic equation is recovered. Yet, in general, due to energy loss, the particle trajectory is subject to an acceleration $J_\perp^{\alpha}$. Clearly, for Eq.~\eqref{PapFinal} to be coordinate invariant, the limit where the size of the particle goes to zero, while keeping $ J_\perp^{\alpha}$ finite, must be taken. Still, the result has the form shown in Eq.~\eqref{PapFinal}, which is the main result of this section and whose consequences are studied in what follows.

\section{Staticity and spherical symmetry}\label{StaticSpherical}

As is well known, the gravitational field produced by the Sun, which is the relevant environment used here to test the method, is properly modeled within the assumptions of staticity and spherical symmetry. The former symmetry means that there exists a timelike Killing vector field that is hypersurface orthogonal \cite{wald1984general}, thus, if $t$ is the time coordinate such that this Killing field is $ \left(\partial/\partial t\right)^a$, there is a $t \to -t$ invariance. Spherical symmetry, on the other hand, can be rigorously defined \cite{wald1984general} and it matches the intuitive notion. Moreover, spherical symmetry gives rise to adapted coordinates in the constant $t$ hypersurfaces that, in the regions under consideration, are a generalization of the conventional spherical coordinates $r$, $\theta$, and $\phi$, with $r>0$.

To proceed, an expression for the spacetime metric is required. Let $f(r)$ and $h(r)$ be such that the static and spherically symmetric metric, in the coordinates at hand, can be brought to the form
\begin{equation}
ds^2 = -f(r)dt^2 + h(r) dr^2 +r^2\left(d\theta^2 + \sin^2\theta d\phi^2\right).
\end{equation}
Importantly, it is assumed that, in the regions of spacetime that are relevant for this work, $f>0 $ and $h>0$, avoiding possible singularities.

In addition, one can verify that any vector field $v^a$ that is invariant under these symmetries must have the following form
\begin{equation}
v^a = A(r) \left(\frac{\partial}{\partial t}\right)^a+ B(r) \left(\frac{\partial}{\partial r}\right)^a,
\end{equation}
where $A(r)$ and $B(r)$ are arbitrary functions. It is assumed that, for simplicity, the nonconservation current $J^a$ is subject to these symmetries, thus, it has two independent components. What is more, the orthogonal current $J_\perp^a$ is completely characterized by a single function $J_\perp^r(r)$ since
\begin{equation}
 J_\perp^t = \frac{h u^r}{f u^t} J_\perp^r,\label{Jperpt}
\end{equation}
where $u^t >0$.

Importantly, an equation of the form of Eq.~\eqref{PapFinal}, when staticity and spherically symmetry are imposed, does not imply that there exists a constant of motion associated with its time component [see for comparison Eq.~\eqref{lcte}]. Still, the trajectory is restricted to a plane, say $\theta = \pi/2$, and there is a constant, along each particle trajectory, associated to the Killing field $\psi^a \equiv (\partial/\partial \phi)^a$. This constant is given by $l \equiv g_{ab}u^a \psi^b = r^2 \sin^2 \theta \dot{\phi} $, where the overdot represents a derivative with respect to the curve parameter. To see that $l$ is constant, one must calculate its change along the trajectory:
\begin{eqnarray}
u^a \nabla_a l &=& u^a \nabla_a\left( u^b \psi_b\right)\nonumber\\
&=& u^b u^a \nabla_{(a}\psi_{b)}+J_\perp^\phi =0,\label{lcte}
\end{eqnarray}
where, in the last step, the Killing equation and the fact that $J_\perp^\phi =0$ are used.

The facts that the motion is restricted to a plane and that there is one constant of motion, make the process of solving Eq.~\eqref{PapFinal} simpler. In addition, it can be seen that the velocity norm, $\kappa\equiv u^a u_a $, is constant along the curve:
\begin{equation}
u^a \nabla_a \kappa =2 u_b u^a\nabla_a u^b=2 u_b J_\perp^b=0.
\end{equation}

Inserting the constants of motion in the velocity normalization condition produces
\begin{equation}
\kappa = g_{ab}u^a u^b=-f \dot{t}^2 + h \dot{r}^2 + \frac{l^2}{r^2}.
\end{equation}
In turn, the last expression can be written as
\begin{eqnarray}
\dot{t}^2 &=&\frac{ h}{f} \dot{r}^2 + \frac{l^2}{fr^2}-\frac{\kappa}{f}.\label{tdot}
\end{eqnarray}
To proceed, note that the $r$ component of Eq.~\eqref{PapFinal} takes the form
\begin{equation}
 \ddot{r}+\frac{f'}{2h} \dot{t}^2 +\frac{h'}{2h} \dot{r}^2 -\frac{r}{h} \dot{\phi}^2 = J_\perp^r ,\label{Jperpr}
\end{equation}
where the prime represents an $r$ derivative. Utilizing Eq.~\eqref{tdot} and the expression for $l$, Eq.~\eqref{Jperpr} can be written as
\begin{equation}
 \ddot{r}+\left(\frac{f'}{f}+ \frac{h'}{h}\right) \frac{\dot{r}^2 }{2}+ \frac{f'}{2fh}\left(\frac{l^2}{r^2}-\kappa\right) -\frac{ l^2}{hr^3} = J_\perp^r .\label{Jperpr1}
\end{equation}

Note that $ \ddot{r}=(1/2) d \dot{r}^2 / dr $, which is a total $r$ derivative. Still, the remaining terms in Eq.~\eqref{Jperpr1} cannot be written as a total $r$ derivative. Thus, in general, Eq.~\eqref{Jperpr1} cannot be trivially integrated. Remarkably, in unimodular gravity it is possible perform this integration. For this reason, the unimodular theory of gravity is studied in what follows.

\section{Unimodular gravity}\label{SecUnimodular}

\subsection{Basic aspects}

The unimodular theory is an alternative theory of gravity with some interesting features \cite{Einstein:1919gv,Anderson:1971pn,vanderBij:1981ym,Buchmuller:1988wx,Unruh:1988in,Henneaux:1989zc,Ng:1990xz,finkelstein2001unimodular,ng2001small,ellis2011trace,PhysRevD.97.084001,Review}. It can be described as a theory in four spacetime dimensions where, in addition to the conventional Einstein-Hilbert action term, there is a Lagrange multiplier that constraints the volume form to coincide with a nondynamical $4$-form. In turn, the presence of this nondynamical $4$-form partially breaks invariance under diffeomorphisms. Concretely, the unimodular action, written as the integral of a scalar density, takes the form
\begin{equation}
S=\int d^4 x\ \frac{1}{2\kappa}\left[\sqrt{-g}R+\lambda\left(\sqrt{-g}-F\right)+\mathcal{L}_M(g,\psi)\right] ,\label{P1}
\end{equation}
where $\lambda$ is the Lagrange multiplier, $F$ is a nondynamical scalar density associated to the nondynamical $4$-form, and $\psi$ collectively represents all the matter fields. The equations of motion are
\begin{eqnarray}
G_{ab}-\frac{1}{2}\lambda g_{ab}&=& \kappa T_{ab}, \label{P9}\\
\sqrt{-g}&=&F,\label{unimodConstraint}\\
\frac{\delta \mathcal{L}_M}{\delta \psi}&=&0,\label{EOMphi}
\end{eqnarray}
where $G_{ab}\equiv R_{ab} - g_{ab}R/2$ is the Einstein tensor and the energy-momentum tensor is defined by Eq.~\eqref{Tab}. Also, Eq.~\eqref{unimodConstraint} is known as the unimodular constraint.

The trace of Eq.~\eqref{P9} generates
\begin{equation}\label{P14}
\lambda=-\frac{1}{2}\left(\kappa T+R\right)
\end{equation}
where $T\equiv g^{ab}T_{ab}$. Using this expression, Eq.~\eqref{P9} can be recast as
\begin{equation}
R^{\rm TL}_{ab}\equiv R_{ab}-\frac{1}{4}R g_{ab}=\kappa\left(T_{ab}-\frac{1}{4}T g_{ab}\right).\label{RicciTL}
\end{equation}
It is easy to see that this equation is traceless, and, together with the unimodular constraint, they comprise a set of ten equations.

The divergence of Eq.~\eqref{RicciTL} produces
\begin{equation}
\frac{1}{4}\nabla_a R =\kappa\left(\nabla^b T_{ab}-\frac{1}{4}\nabla_a T \right).
\end{equation}
Therefore, in cases when $\nabla^b T_{ab}=0$, $\tilde{\Lambda}\equiv (R +\kappa T)/4 $ is a constant. Interestingly, inserting $\tilde{\Lambda}$ into Eq.~\eqref{RicciTL} yields the conventional Einstein equations with $\tilde{\Lambda}$ acting as a cosmological constant. In other words, unimodular gravity, with the additional assumption that the energy-momentum tensor is divergence free, is dynamically equivalent to GR with a cosmological constant that arises as an integration constant. As such, this constant is completely independent from the vacuum energy of the matter fields, offering a plausible explanation to the ``cosmological constant problem'' \cite{Weinberg} (see also Ref.~\onlinecite{Bengochea2020}).

Moreover, $F$, being nondynamical, does not transform under diffeomorphisms. As a consequence, the action \eqref{P1} is not invariant under all diffeomorphisms. To show this, consider the simpler case when $T_{ab}=0$. The action variation under an infinitesimal diffeomorphism associated with $\xi^a$ takes the form
\begin{equation}\label{d5}
\delta S =\int d^4x\ \sqrt{-g} \lambda\nabla_a\left(\frac{F}{\sqrt{-g}}\xi^a\right),
\end{equation}
where the Bianchi identity $\nabla_a G^{ab}=0$ is used. On shell, $F=\sqrt{-g}$, and the action is invariant if and only if $\nabla_a\xi^a=0$. Now, to find the conservation law, it is necessary to express a divergence free vector field $\xi^a$ in terms of an arbitrary tensor. This is achieved through an antisymmetric tensor $\alpha_{ab}$. Let
\begin{equation}
 \xi^a = \epsilon^{abcd}\nabla_b\alpha_{cd},
 \end{equation}
where $\epsilon_{abcd}$ is the metric volume form, i.e., the $4$-form such that $\epsilon^{abcd}\epsilon_{abcd}=-4!$ and $\nabla_a \epsilon_{bcde}=0$. Then,
\begin{eqnarray}
\nabla_a \xi^a&=&\nabla_a\left(\epsilon^{abcd}\nabla_b\alpha_{cd}\right) \nonumber\\
&=&\frac{1}{2}\epsilon^{abcd}\left(\nabla_a\nabla_b-\nabla_b\nabla_a\right) \alpha_{cd} \nonumber\\
&=&\epsilon^{abcd}{R_{[abc]}}^e\alpha_{ed} =0,
\end{eqnarray}
where the squared brackets denote the totally antisymmetric part and, in the last step, the identity ${R_{[abc]}}^d=0$ is used.

The matter action variation is still given by Eq.~\eqref{actionvariation} but now the vector field $\xi^a$ is restricted to be divergence free. Expressing $\xi^a$ in terms of $\alpha_{ab}$ yields
\begin{equation}
0=\delta S_M = \int d^4x\ \sqrt{-g} \alpha_{de} \epsilon^{bcde} \nabla_c \nabla^a T_{ab} ,
\end{equation}
where a double integration by parts is performed. Given that $\alpha_{ab}$ is arbitrary, the corresponding energy nonconservation law can be readily found; it reads
\begin{equation}
0= \epsilon^{bcde} \nabla_c \nabla^a T_{ab}.\label{ConservLawUG}
\end{equation}
Clearly, this conservation law is satisfied when $ \nabla^a T_{ab}=0$, in which case the theory reduces to GR, as is argued above. Yet, Eq.~\eqref{ConservLawUG} has more general solutions. In fact, denoting $j^a \equiv \nabla_b T^{ab}$, Eq.~\eqref{ConservLawUG} implies that the exterior derivative of $j_a$ has to vanish, namely ${\rm d}j=0$. What is more, using Poincare's lemma \cite{Nakahara:2016}, it is possible to show that, if spacetime is contractible, as it is assumed hereon, there exists a scalar function $\Phi$, called the nonconservation potential, such that
\begin{equation}
j_a= -\nabla_a \Phi.\label{Poincare}
\end{equation}
This fact makes unimodular gravity tractable with the standard particle trajectory methods. The other result is a generalization a well-known theorem, which is presented in the next subsection.

\subsection{Generalized Birkhoff theorem}

The Birkhoff theorem states that the only spherically symmetric solution of the vacuum Einstein equation, $R_{ab}=0$, is the Schwarzschild solution, which is static. In this subsection, an analogous result is derived in the context of vacuum unimodular gravity, following the method presented in Ref. \cite[page 468]{Hartle}. The goal is to solve Eq.~\eqref{RicciTL} in vacuum, namely, $R^{\rm TL}_{ab} =0$, in spherical symmetry. In adapted coordinates, the most general spherically symmetric line element, which is not necessarily static, is
\begin{eqnarray}
ds^2&=&-A\left(t,r\right)dt^2+2B\left(t,r\right)dtd r+C\left(t,r\right)dr^2\nonumber\\
&&+r^2\left(d\theta^2+r^2\sin^2\theta\mathrm{d }\phi^2\right),\label{P40}
\end{eqnarray}
where $A$, $B$, and $C$ are arbitrary functions. With a coordinate transformation, $\tilde{t}=\tilde{t}(t,r)$, it is possible to absorb the $B$ function, obtaining
\begin{equation}
ds^2=-e^{\nu\left(\tilde{t},r\right)}d\tilde{t}^2+ e^{\rho \left( \tilde{t},r\right)}dr^2+r^2\left(d\theta^2+\sin^2\theta \ d\phi^2\right ).\label{P41}
\end{equation}

With this metric, the $\tilde{t}r$ component of $R^{\rm TL}_{ab} =0$ is simply
\begin{equation}
\frac{1}{r}\frac{\partial \rho}{\partial \tilde{t}}=0,\label{RTLtr}
\end{equation}
which is solved by $\rho=\rho(r)$. In addition,
\begin{equation}
0=r e^{\rho-\nu}R^{\rm TL}_{\tilde{t}\tilde{t}} +r R^{\rm TL}_{rr}=\nu'+ \rho',\label{RTLtt+rr}
\end{equation}
which implies
\begin{equation}
\nu\left(\tilde{t},r\right)=\beta\left(\tilde{t}\right)-\rho\left(r\right).\label{P49}
\end{equation}
What is more, the $\theta\theta$ component of $R^{\rm TL}_{ab} =0$ becomes
\begin{equation}\label{P50}
0=\frac{4 }{r^2}e^{\rho} R^{\rm TL}_{\theta\theta} =\frac{2}{r^2} (e^{\rho}-1)+\rho'^2-\rho'',
\end{equation}
whose solution is
\begin{equation}\label{P51}
\rho\left(r\right)=-\ln\left(1- \frac{2M}{r} +\Lambda r^2\right),
\end{equation}
with $M$ and $\Lambda$ constants known as the mass and the cosmological constant, respectively.

Using Eqs.~\eqref{P49} and \eqref{P51}, the metric \eqref{P41} takes the form
\begin{eqnarray}
ds^2&=& -e^{\beta\left(\tilde{t}\right)}\left(1- \frac{2M}{r} +\Lambda r^2\right)d \tilde{t}^2 +\frac{dr^2}{1- \frac{2M}{r} +\Lambda r^2}\nonumber\\
&&+r^2\left(d\theta^2+\sin^2\theta \ d\phi^2\right).\label{P52}
\end{eqnarray}
Finally, it is possible to change coordinates once again and introduce $\hat{t}=\hat{t}(\tilde{t})$ to absorb $\beta$, producing
\begin{eqnarray}\label{P53}
ds^2&=&-\left(1- \frac{2M}{r} +\Lambda r^2\right)d \hat{t}^2+\frac{dr^2}{1- \frac{2M}{r} +\Lambda r^2}\nonumber\\
&&+r^2\left(d\theta^2+\sin^2\theta \ d\phi^2\right),\label{BirkhoffFinal}
\end{eqnarray}
which is the Schwarzschild, Schwarzschild-de Sitter, or Schwarzschild-antide Sitter metric, depending on the sign of $\Lambda$, which are static. In addition, it can be verified that the metric \eqref{P53} solves all the components of $R^{\rm TL}_{ab} =0$. This result is clearly compatible with the fact, discussed above, that, in vacuum, unimodular gravity reduces to GR with a cosmological constant. In this sense, this result is a generalization, relevant for unimodular gravity, of Birkhoff's celebrated theorem.

\section{Empirical constraints}\label{Empirical}

In this section, Solar system observations are used to test particular models for energy nonconservation. This is done as a proof of concept; the goal is to show that energy nonconservation produces physical effects. For simplicity, a situation where all the calculations can be performed analytically is chosen. The proposal is to consider the vacuum propagation of pointlike \emph{test} particles that lose energy according to unimodular gravity, in a background that is a solution of this theory. Also, spherical symmetry is assumed, for the background metric and for the energy nonconservation potential, since it is a good approximation to account for Solar system observations. Moreover, $c$ is restored to be able to expand in $c^{-2}$. 

Interestingly, vacuum unimodular gravity, under the symmetries at hand, is a theory where Eq.~\eqref{Jperpr1} is a total derivative, which greatly simplifies the calculations. This follows from Eq.~\eqref{Poincare} and from the generalized Birkhoff theorem, which implies $f = h^{-1}$. As a consequence, the following identities can be verified:
\begin{eqnarray}
\frac{f'}{f}+\frac{h'}{h}&=& 0,\\
 \frac{f'}{2fh}\left(\frac{l^2}{r^2}-\kappa\right) -\frac{ l^2}{hr^3} &=&\frac{d}{dr}\left[\frac{f}{2}\left(\frac{l^2}{r^2}-\kappa\right) \right].
\end{eqnarray}
Now, Eq.~\eqref{Poincare} states that $j_a =-\nabla_a\Phi/c^2$, where the factor $c^{-2}$ is introduced so that the loss of energy is at the same order than the gravitational effects. Clearly, in spherical symmetry, $\Phi=\Phi(r)$, and thus, the right-hand side of Eq.~\eqref{Jperpr1} is also a total $r$ derivative. Notice that the right-hand side of Eq.~\eqref{Jperpr1}, when given by this derivative, should be regarded as an additional, albeit well motivated, assumption, since what appears on the right-hand side of Eq.~\eqref{PapFinal} is proportional to the integral of $j_a$ in the limit when the particle's size goes to zero.

Eq.~\eqref{Jperpr1}, with these considerations, becomes
\begin{equation}
\frac{1}{2} \frac{d\dot{r}^2}{dr}+ \frac{d}{dr}\left[\frac{f}{2}\left(\frac{l^2}{r^2}-\kappa\right) \right]=-\frac{1}{c^2} \frac{d \Phi}{dr},
\end{equation}
which  can be integrated to produce
\begin{equation}
E=\frac{1}{2} \dot{r}^2 +V_{\rm eff}(r) ,\label{AnalogEnerg}
\end{equation}
where $E$ is an integration constant and 
\begin{equation}
V_{\rm eff}(r)\equiv \frac{f}{2}\left(\frac{l^2}{r^2}-\kappa\right) +\frac{1}{c^2}\Phi. \label{Veff}
\end{equation}
Clearly, Eq.~\eqref{AnalogEnerg} has the form of an energy equation in conventional Newtonian dynamics for a system of one degree of freedom.

Before proceeding, note that the observed  \cite{PlanckCollaboration} cosmological constant, $\Lambda$, is such that, for all the relevant values of $r$, $\Lambda r^2 \ll 1$. Thus, such contributions are neglected and $f=1-2M/(c^2 r)$. In consequence, the utilized metric is the Schwarzschild metric and the only unconventional effects that are considered here are due to $\Phi$. In what follows, comparisons with Solar system data are performed.

\subsection{Null trajectories}

Given that light follows null trajectories contained in a plane, $\kappa=0$ and $\theta=\pi/2$. What is more, it is useful to define $b\equiv l/\sqrt{2E}$, which, when $\Phi=0$, is the impact parameter. Assuming\footnote{The case where $l<0$ is related to what is done here by changing the coordinates' orientation; radial trajectories, where $l=0$, must be analyzed independently.} that $l> 0$, it is possible to express Eq.~\eqref{AnalogEnerg} as
\begin{equation}
\frac{1}{b^2}= \frac{1}{l^2}\dot{r}^2 + V_{\rm eff}^{\rm n},\label{AnalogEnergNull}
\end{equation}
where
\begin{equation}
V_{\rm eff}^{\rm n}\equiv \frac{1}{r^2}f+\frac{2}{c^2l^2}\Phi_{\rm n},\label{Veffn}
\end{equation}
and $\Phi_{\rm n}$ is the nonconservation potential for null trajectories.

\begin{figure}
\begin{center}
 \includegraphics[width=0.5\textwidth]{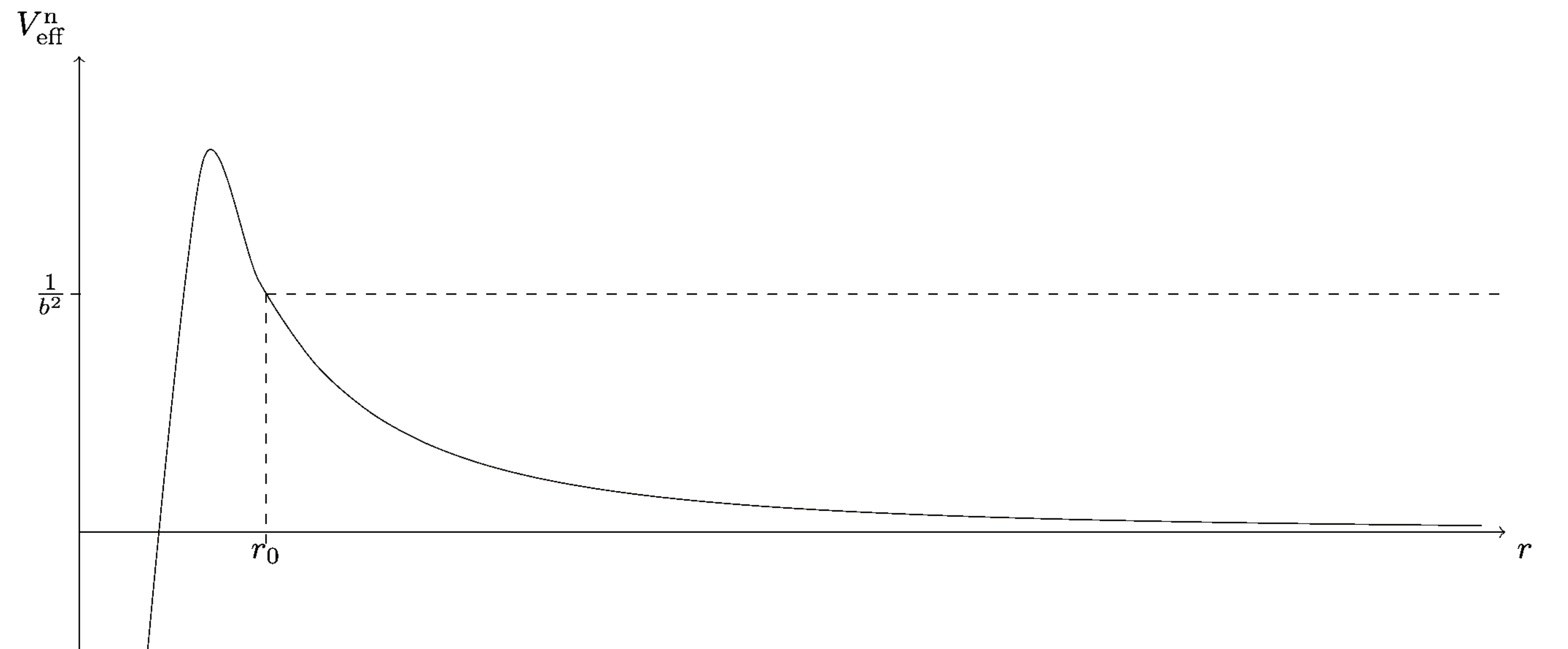} \caption{\label{FiguraVeffN}Effective potential for null trajectories as a function of the radial coordinate $r$. For a given $b$, there is a radius of maximal approach, $r_0$.}
 \end{center}
\end{figure}

To proceed, a concrete expression for $\Phi_{\rm n}$ is necessary. Such a potential is assumed to vanish as $r\to \infty$ to avoid sourcing energy nonconservation by objects located arbitrarily far away. Moreover, $\Phi_{\rm n}$ is taken to be such that it maintains the structure of the maxima and minima of the effective potential as compared with the $\Phi_{\rm n}=0$ case. This is because a change in this structure would produce dramatic effects that have not been observed. Given that $V_{\rm eff}^{\rm n}$, when $\Phi_{\rm n}=0$, has $r^{-2}$ and $r^{-3}$ terms, the following form for the nonconservation potential is considered:
\begin{equation}
\Phi_{\rm n}=\frac{b^{\rm n}_2}{r^2}+\frac{b^{\rm n}_3}{r^3},
\end{equation}
where $b^{\rm n}_2$ and $b^{\rm n}_3$ are free parameters. From inspection, it can be verified that, for these parameters to keep the form of the effective potential, it is necessary that
\begin{eqnarray}
l^2c^2 +2b^{\rm n}_2 >0,\label{condnul1}\\
Ml^2-b^{\rm n}_3 >0.\label{condnul2}
\end{eqnarray}
These are restrictions on $l$ and on the nonconservation parameters, and are assumed hereon. The qualitative form of $V_{\rm eff}^{\rm n}$ is plotted in Fig.~\ref{FiguraVeffN}. Importantly, in general, the parameters of this nonconservation potential cannot be absorbed into $M$ and $l$.

\subsubsection{Deflection of light}

The deflection of light was the first prediction of GR to be empirically confirmed, and refinements of this effect have been proposed, arising, for example, from QED radiative corrections \cite{Drummond} or from higher-order geometrical theories \cite{Accioly}. Here, this effect is computed for unimodular gravity and with the above presented nonconservation potential.

Light trajectories are described by the angle $\phi$ as a function of $r$. In turn, $\phi(r)$ can be found after integrating $d\phi/dr = \dot{\phi}/\dot{r}$; the numerator is $\dot{\phi}=l/r^2$, and the denominator is obtained from Eq.~\eqref{AnalogEnerg}, producing
\begin{equation}
\frac{d\phi}{dr}= \frac{1}{r^2\sqrt{\frac{1}{b^2}-\frac{ f}{r^2} -\frac{2}{l^2 c^2}\Phi_{\rm n}}}.\label{dphidrluz}
\end{equation}
To calculate the total deflection angle, $\Delta \phi_{\rm def}$, it is customary to consider that the light beam emerges and returns to $r \to \infty$. Given the symmetries of the problem, the net effect is twice the $r$-integral of Eq.~\eqref{dphidrluz} from infinity to the radius of maximal approach, $r_0$, which corresponds to the radius where $\dot{r}=0$. From Eq.~\eqref{AnalogEnergNull} at $r_0$, it can be seen that
\begin{equation}
\frac{1}{b^2}= \frac{1}{r_0^2}f(r_0)+\frac{2}{c^2l^2}\Phi_{\rm n}(r_0).\label{bvsr0}
\end{equation}
With this, it can be verified that the total deflection angle satisfies
\begin{equation}
\Delta \phi_{\rm def} = \int_{r_0}^{\infty}\frac{2\ dr}{r^2\sqrt{ \frac{f(r_0)}{r_0^2}-\frac{ f(r)}{r^2}+\frac{2}{c^2l^2}\left[\Phi_{\rm n}(r_0) -\Phi_{\rm n}(r)\right]}}.\label{integraldef2}
\end{equation}

To integrate, it is helpful to expand the integrand in powers of $c^{-2}$; the leading contribution is
\begin{equation}
\Delta \phi^{(0)}_{\rm def} = \int_{r_0}^{\infty}\frac{2\ dr}{r^2\sqrt{ \frac{1}{r_0^2}-\frac{ 1}{r^2}}}=\pi .\label{pi}
\end{equation}
Moreover, the first correction takes the form
\begin{eqnarray}
\Delta \phi^{(1)}_{\rm def}&=&\frac{1}{c^2} \int_{r_0}^{\infty}\frac{dr \left\{ \frac{2M}{r_0^3}-\frac{2M }{r^3}-\frac{2}{l^2}\left[\Phi_{\rm n}(r_0) -\Phi_{\rm n}(r)\right]\right\}}{r^2\left(\frac{1}{r_0^2}-\frac{ 1}{r^2}\right)^{3/2}}\nonumber\\
&=&\frac{4(Ml^2-b^{\rm n}_3)}{c^2l^2r_0} - \frac{\pi b^{\rm n}_2}{c^2l^2} .
\end{eqnarray}
Considering that $\Delta \phi_{\rm def} =\pi$ corresponds to a light beam that follows a straight trajectory, the relevant deflection angle is
\begin{equation}
\delta \phi_{\rm def}\equiv\Delta \phi_{\rm def} - \pi = \frac{4(Ml^2-b^{\rm n}_3)}{c^2l^2r_0} - \frac{\pi b^{\rm n}_2}{c^2l^2} .
\end{equation}

At this point, it is convenient to use $r_0 = b+O\left(c^{-2}\right)$, which is derived from Eq.~\eqref{bvsr0}. Also, Eq.~\eqref{AnalogEnergNull} in the region $r\to \infty$ implies $l = b$, where the fact that $\dot{r}\to 1$ is utilized\footnote{Here, $b$ is the geometrical impact parameter up to $O(c^{-2})$ corrections, which can be neglected.}. The final expression reads
\begin{equation}
\delta \phi_{\rm def} =\frac{4M}{c^2b} - \frac{\pi b^{\rm n}_2}{c^2b^2} -\frac{4b^{\rm n}_3}{c^2b^3}.\label{deltaphi}
\end{equation}
As expected, $\delta \phi_{\rm def} $ has the GR term plus modifications due to $b^{\rm n}_2$ and $b^{\rm n}_3$. The former produces an effect that goes like $b^{-2}$, while the latter goes like $b^{-3}$, in agreement with naive dimensional considerations.

Bounds on $b^{\rm n}_2$ and $b^{\rm n}_3$ are set by using the well-known Parameterized Post-Newtonian formalism (PPN) \cite{WillBook}. In turn, the PPN bounds are set using light rays that pass extremely close to the surface of the Sun, therefore, the impact parameter is approximately the Sun's radius, $R_\odot$. Moreover, the corresponding PPN expression reads \cite{Will2006}
\begin{equation}
 \delta \phi^{\rm PPN}_{\rm def}= (1+\gamma) \frac{2 M_\odot}{c^{2} R_\odot},
\end{equation}
where $\gamma$ is the PPN parameter and $M_\odot$ is the mass of the Sun. Comparison with Eq.~\eqref{deltaphi} yields
\begin{equation}
 \frac{\pi b^{\rm n}_2}{2 M_\odot R_\odot} +\frac{2b^{\rm n}_3}{ M_\odot R_\odot^2}= 1-\gamma .
\end{equation}
The corresponding PPN bounds are \cite{refId0}
\begin{equation}
-2.0 \times 10^{-4} < \gamma -1 < 0.4 \times 10^{-4},
\end{equation}
which, using $R_\odot \approx 5 \times 10^5 M_\odot$, translate to
\begin{equation}
-51 < 4.0 \frac{b^{\rm n}_2}{M_\odot^2 } + (1.1\times 10^{-5})\frac{ b^{\rm n}_3}{M_\odot^3} < 255.\label{boundslightdef}
\end{equation}
Note that a change in the reference scale, form $M_\odot$ to, say, ${\rm GeV}\sim 10^{-57} M_\odot$, would drastically change the numbers in Eq.~\eqref{boundslightdef}. Finally, it is important to emphasize that these limits are compatible with conditions \eqref{condnul1}-\eqref{condnul2} provided that
\begin{equation}
\frac{ b^{\rm n}_3}{M_\odot^3} <2.5 \times 10^{11}.
\end{equation}

An interesting proposal\footnote{Due to an anonymous referee.} is to find an effective index of refraction, $n_{\rm eff}$, that describes the effects of gravity and energy nonconservation. In Ref.~\cite{Fischbach} a gravitational effective index of refraction is found by comparing the effects of a media and those of spacetime curvature in Maxwell's equations. Thus, to utilize the methods of Ref.~\cite{Fischbach}, it is necessary to incorporate energy nonconservation into an effective metric. For the case at hand, this can be done by inspecting Eq.~\eqref{Veffn} and noticing that there is an effective function, $f_{\rm eff}$, such that $V_{\rm eff}^{\rm n}$ has the same functional form as in GR but with $f_{\rm eff}$ in place of $f$. This function has the form
\begin{equation}
 f_{\rm eff} \equiv f+\frac{2r^2}{c^2l^2}\Phi_{\rm n}=1+\frac{2b^{\rm n}_2}{c^2l^2}  -\frac{2M }{c^2 r}\left(1-\frac{b^{\rm n}_3 }{Ml^2}\right),
\end{equation}
and the corresponding ``effective metric'' has the form of Eq.~\eqref{BirkhoffFinal} but with $f_{\rm eff}$ instead of $f $. Note, however, that energy nonconservation is not ``geometrizable'' in the sense that $f_{\rm eff}$ depends on the light trajectory via $l$. With this observation, finding $n_{\rm eff}$ reduces to writing the metric in isotropic coordinates and reading off the time and spatial components~\cite{Fischbach}; the result is
\begin{equation}
n_{\rm eff} =1-\frac{b^{\rm n}_2}{c^2 l^2}+\frac{ M -b^{\rm n}_3 /l^2 }{c^2  r},\label{neff}
\end{equation}
where, for simplicity, an expansion on $c^{-2}$ is performed. Eq.~\eqref{neff} reduces to the correct results when the parameters for energy nonconservation are set to zero. What is more, light trajectories could be found from $n_{\rm eff}$ following the iterative procedure described in Ref.~\cite{Fischbach}. More interestingly, Eq.~\eqref{neff} could lead to new experiments to search for static and spherically symmetric forms of energy nonconservation, particularly in the realm of tabletop optics experiments.

\subsection{Time delay}

The time delay, as originally devised by Shapiro \cite{Shapiro}, is calculated in this subsection for the case where there is energy nonconservation. This delay concerns the time spend by a light signal that travels from the Earth to a reflecting satellite that orbits the Sun, and back, when the light beam passes close to the Sun. In addition, this effect is calculated neglecting the change in the position of the Earth and the satellite. The relevant expression is $t=t(r)$, which is obtained by integrating $dt/dr = \dot{t}/\dot{r}$, where the fact that $\dot{r}\neq 0$ is used. From Eq.~\eqref{tdot}, it is possible to write
\begin{equation}
\frac{dt}{dr}=\pm\frac{1}{cf} \sqrt{1+\frac{l^2f}{ r^2 \dot{r}^2}}\ ,\label{dtdr}
\end{equation}
where the sign has to be appropriately chosen depending on whether $r$ grows or decreases. Eq.~\eqref{AnalogEnergNull} evaluated at an arbitrary radius $r$ minus the same equation at $r_0$ is 
\begin{equation}
\frac{1}{l^2}\dot{r}^2 = \frac{1}{r_0^2}f(r_0)- \frac{1}{r^2}f(r) +\frac{2}{c^2l^2}\left[\Phi_{\rm n}(r_0)-\Phi_{\rm n}(r)\right].\label{AnalogEnergNull0}
\end{equation}
In turn, this can be used to bring Eq.~\eqref{dtdr} to the form
\begin{widetext}
\begin{equation}
\frac{dt}{dr}=\pm\frac{1}{cf} \sqrt{1+\frac{f/r^2}{ \frac{1}{r_0^2}f(r_0)- \frac{1}{r^2}f(r) +\frac{2}{c^2l^2}\left[\Phi_{\rm n}(r_0)-\Phi_{\rm n}(r)\right]}}\ .
\end{equation}

The time it takes the light to travel from $r_1$ to $r_2$ is obtained through integration:
\begin{equation}
\Delta t(r_1,r_2)=\int_{r_1}^{r_2} dr \frac{dt}{dr} =\pm\int_{r_1}^{r_2} \frac{dr}{cf} \sqrt{1+\frac{f/r^2}{ \frac{1}{r_0^2}f(r_0)- \frac{1}{r^2}f(r) +\frac{2}{c^2l^2}\left[\Phi_{\rm n}(r_0)-\Phi_{\rm n}(r)\right]}}.
\end{equation}
Note that $\Delta t(r_1,r_2)=\Delta t(r_2,r_1)$ as the sign that arises when changing the limits of the integral compensates the sign associated with $\dot{r}$. This is also clear on physical grounds (recall that the change of position of the Earth and the satellite is neglected). The total time traveled by the light signal is given by $\Delta t_{\rm tot} \equiv 2 \Delta t(r_0,R_E)+ 2 \Delta t(r_0,R_S)$, where $R_E$ and $R_S$ are, the radius of the orbit of the Earth and the satellite, respectively. Thus, it is necessary to find expressions for $\Delta t(r_0,R)$ for an arbitrary $R>r_0$.

Again, the integral can be solved analytically when expanding in $c^{-2}$. To lowest order, the integral becomes
\begin{equation}
\Delta t(r_0,R)^{(0)}=\frac{1}{c}\int_{r_0}^{R} dr \frac{r}{ \sqrt{r^2 -r_0^2}}=\frac{1}{c}\sqrt{R^2-r_0^2}.
\end{equation}
The next order effect is
\begin{eqnarray}
\Delta t(r_0,R)^{(1)}&=&\int_{r_0}^{R} \frac{l^2 M (r-r_0) (2 r+3 r_0)+r r_0^4 \left[\Phi_{\rm n} (r)-\Phi_{\rm n} (r_0)\right]}{c^3 l^2 \left(r^2-r_0^2\right)^{3/2}}\nonumber\\
&=&\frac{M }{c^3} \left[\sqrt{\frac{R-r_0}{R+r_0}}+4\arcsinh \left(\sqrt{\frac{R-r_0}{2r_0}}\right)\right]\nonumber\\
&&+\frac{-\pi b^{\rm n}_2 R r_0+4 b^{\rm n}_2 R r_0 \arctan\left(\frac{R-\sqrt{R^2-r_0^2}}{r_0}\right)-2 b^{\rm n}_3 (2 R+r_0) \sqrt{\frac{2 R}{R+r_0}-1}}{2 c^3 l^2 R}.
\end{eqnarray}
The ``excess time,'' $\delta t $, is defined as $\Delta t_{\rm tot} $ minus the flight time in flat spacetime and in a theory with energy conservation (i.e., when $M$, $b^{\rm n}_2$, and $b^{\rm n}_3$ vanish). It is given by
\begin{eqnarray}
\delta t &=& 2 \Delta t(r_0,R_E)+ 2 \Delta t(r_0,R_S)- \frac{2}{c}\sqrt{R_E^2-r_0^2}- \frac{2}{c}\sqrt{R_S^2-r_0^2}\nonumber\\
&=&\frac{-\pi b^{\rm n}_2 R_E r_0+4 b^{\rm n}_2 R_E r_0 \arctan\left(\frac{R_E-\sqrt{R_E^2-r_0^2}}{r_0}\right)-2 b^{\rm n}_3 (2 R_E+r_0) \sqrt{\frac{2 R_E}{R_E+r_0}-1}}{c^3 l^2 R_E} \nonumber\\
&&+ \frac{-\pi b^{\rm n}_2 R_S r_0+4 b^{\rm n}_2 R_S r_0 \arctan\left(\frac{R_S-\sqrt{R_S^2-r_0^2}}{r_0}\right)-2 b^{\rm n}_3 (2 R_S+r_0) \sqrt{\frac{2 R_S}{R_S+r_0}-1}}{c^3 l^2 R_S}\nonumber\\
&\approx& \frac{ 4M}{c^3}\left[1+ \ln \left(\frac{4 R_E R_S}{r_0^2}\right)\right] - \frac{4 b^{\rm n}_3}{c^3 r_0^2},\label{delta t}
 \end{eqnarray}
\end{widetext}
where, in the last step, an expansion in $r_0/R_E $ and $r_0/R_S$ is performed. Also, from Eq.~\eqref{AnalogEnergNull0} in the $r\to \infty$ limit, it is possible to show that $l = r_0 + O\left(c^{-2}\right)$, which is used in the last term of Eq.~\eqref{delta t}.

Importantly, it can be verified that the first part of Eq.~\eqref{delta t} coincides with the corresponding GR expression (see, e.g., \cite[page 214]{Hartle}). Moreover, $b^{\rm n}_3$ produces a shift on $\delta t$ that is independent on $R_E $ and $R_S$, and, at this level, there are no effects due to $b^{\rm n}_2$. Unfortunately, when analyzing this type of empirical data, a constant shift in $\delta t$ is usually discarded. In fact, the PPN limits \cite{Will2006} are set using the logarithmic part of $\delta t$. Therefore, PPN limits set with time delay do not lead to limits on $b^{\rm n}_3$, which was impossible to foresee before making the calculations.

\subsection{Timelike trajectories}

Under the symmetries at hand, timelike trajectories, which describe the propagation of massive pointlike particles, are also contained in the $\theta=\pi/2$ plane, but now $\kappa=-c^2$. Therefore,
\begin{equation}
E-\frac{c^2}{2}=\frac{1}{2} \dot{r}^2 +V^{\rm t}_{\rm eff},\label{AnalogEnergTimelike}
\end{equation}
where the superindex ${\rm t}$ refers to timelike curves. The effective potential is
\begin{eqnarray}
V^{\rm t}_{\rm eff}&=& \frac{f}{2}\left(\frac{l^2}{r^2}+c^2\right)-\frac{c^2}{2} +\frac{1}{c^2}\Phi_{\rm t} \nonumber\\
&=&-\frac{l^2 M}{c^2r^3}+ \frac{l^2}{2r^2} -\frac{ M}{r } +\frac{1}{c^2}\Phi_{\rm t} .
\end{eqnarray}
Here, $\Phi_{\rm t}=\Phi_{\rm t}(r)$ is the nonconservation potential for timelike trajectories. Also, note that a term $-c^2/2$ is introduced in both sides of Eq.~\eqref{AnalogEnergTimelike} so that the energy conserving part of $V^{\rm t}_{\rm eff}$ tends to zero as $r\to \infty$.

\begin{figure}
\begin{center}
 \includegraphics[width=0.5\textwidth]{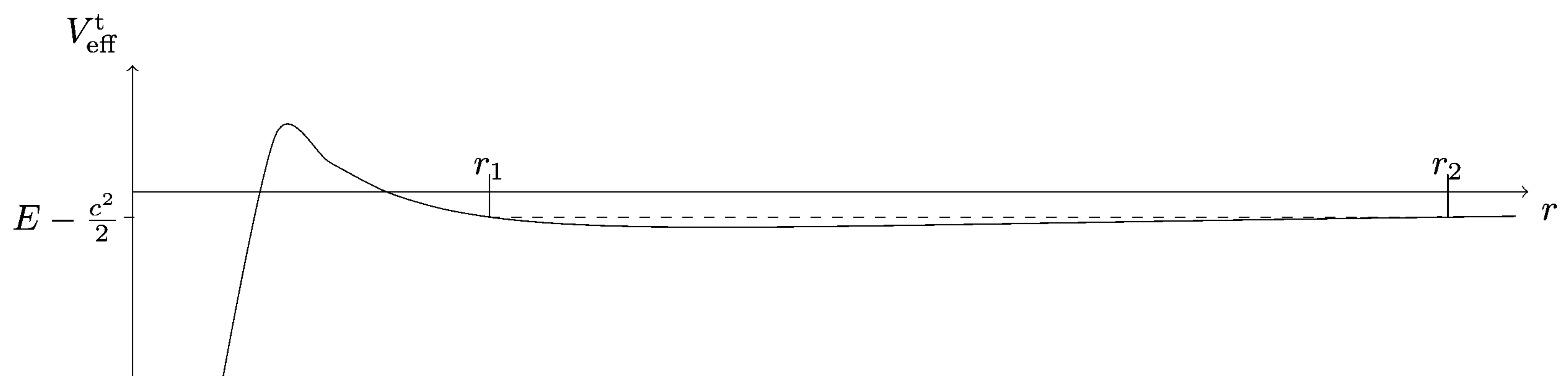}
 \caption{\label{FiguraVeffT}$V^{\rm t}_{\rm eff}$ as a function of $r$. When $E<c^2/2$, the potential admits orbits that are confined between $r_1$ and $r_2$.}
 \end{center}
 \end{figure}

The effective potential $V^{\rm t}_{\rm eff}$, without energy nonconservation, has terms that are proportional to $r^{-1}$, $r^{-2}$, and $r^{-3}$. Therefore, for $\Phi_{\rm t}$ to maintain the form of $V^{\rm t}_{\rm eff}$, it is necessary that
\begin{equation}
\Phi_{\rm t}=\frac{b^{\rm t}_1}{r}+\frac{b^{\rm t}_2}{r^2}+\frac{b^{\rm t}_3}{r^3},
\end{equation}
where the free parameters, in this case, are given by $b^{\rm t}_1$, $b^{\rm t}_2$, and $b^{\rm t}_3$. Moreover, it is necessary that the coefficients of $r^{-3}$, $r^{-2}$, and $r^{-1}$, are, respectively, negative, positive, and negative, and, for the effective potential to have two extrema, it is also needed that
\begin{eqnarray}
Mc^2 - b^{\rm t}_1 &>& 0,\label{condt1}\\
c^2 l^2+2b^{\rm t}_2 &>&0 ,\\
 M l^2 - b^{\rm t}_3 &>&0, 
 \end{eqnarray}
 and, in addition,
 \begin{equation}
12 \left( Mc^2- b^{\rm t}_1\right)\left( Ml^2 - b^{\rm t}_3\right) \leq \left(l^2c^2 +2b^{\rm t}_2\right)^2. \label{condt4}
\end{equation}
The behavior of $V^{\rm t}_{\rm eff}$ is plotted in Fig.~\ref{FiguraVeffT}. Note that $V^{\rm t}_{\rm eff}$ allows the particles to follow orbits, that is, timelike trajectories that are contained in between two radii $r_1$ and $r_2 \geq r_1$; this happens when $E$ is such that  $V_{\rm eff}(r_{\rm min}) \leq E-c^2/2 <0 $, where $r_{\rm min}$ is the radius that minimizes the effective potential. These additional conditions, together with Eqs.~\eqref{condt1}-\eqref{condt4}, are assumed in what follows. Again, it can be verified that the free parameters cannot be all absorbed into effective values for $M$ and $l$.

\subsubsection{Perihelion shift}

Orbits can be described by the angular coordinate as a function of the radial coordinate: $\phi=\phi(r)$. Once again, this function can be obtained by integrating $d\phi/dr = \dot{\phi}/\dot{r}$. From the definition of $l$ and Eq.~\eqref{AnalogEnergTimelike}, it is possible to write
\begin{equation}
\frac{d\phi}{d r}=\pm \frac{l/r^2}{\sqrt{2E - c^2 -2V^{\rm t}_{\rm eff}(r)}},
\end{equation}
where the sign is positive (negative) whenever $\dot{r}$ is positive (negative). Let $r_1>0$ and $r_2 > r_1$ be the radii where $\dot{r}=0$, which are known as the return points of the orbit\footnote{There are three solutions to $\dot{r}=0$, however, one solution does not correspond to orbits. In addition, circular orbits ($r_1=r_2$) can be considered by taking, at the end of the calculation, the $\epsilon \to 0$ limit, where $\epsilon$ is the orbit's eccentricity.}. From Eq.~\eqref{AnalogEnergTimelike}, it follows that
\begin{equation}
E-\frac{c^2}{2}=V^{\rm t}_{\rm eff}(r_1) ,\label{Vret}
\end{equation}
and
\begin{equation}
l= \sqrt{\frac{c^2[f(r_2)-f(r_1)]+\frac{2}{c^2}[\Phi_{\rm t}(r_2)-\Phi_{\rm t}(r_1)]}{\frac{f(r_1)}{r_1^2}-\frac{f(r_2)}{r_2^2}}},\label{lprec}
\end{equation}
where $l$ is taken as positive.

An orbit is defined as the trajectory from the minimal radius $r_1$, to $r_2>r_1$, and back. It turns out that the angle spanned when going from $r_1$ to $r_2$ is equal to that from $r_2$ to $r_1$, since the change in the integration limits compensates the global sign. Thus, the angle spanned during a full orbit is
\begin{eqnarray}
\Delta \phi &=& 2l \int_{r_1}^{r_2}\frac{dr}{r^2 \sqrt{2V^{\rm t}_{\rm eff}(r_1) -2V^{\rm t}_{\rm eff}(r)}},
\end{eqnarray}
where Eq.~\eqref{Vret} is used. Moreover, Eq.~\eqref{lprec} can be utilized to replace $l$, which appears inside $V^{\rm t}_{\rm eff}$, by a function of $r_1$ and $r_2$.

To integrate, an expansion in powers of $c^{-2}$ is performed. The dominant term is
\begin{eqnarray}
\Delta \phi^{(0)} &=& 2\int_{r_1}^{r_2} \frac{dr\ \sqrt{r_1 r_2}}{r \sqrt{(r-r_1) (r_2-r)}}=2\pi,
\end{eqnarray}
and the next order effect is given by
\begin{eqnarray}
\Delta \phi^{(1)} &=& \int_{r_1}^{r_2} \left\{\frac{2 M [r (r_1+r_2)+r_1 r_2]}
{c^2 r^2 \sqrt{r_1 r_2} (r-r_1)^{1/2} (r_2-r)^{1/2}}\right. \nonumber\\
&& -\frac{\sqrt{r_1 r_2} (r_1+r_2)}{c^2 M } \frac{r \Phi_{\rm t} (r)}
{ (r-r_1)^{3/2} (r_2-r)^{3/2}} \nonumber\\
&&+\frac{\sqrt{r_1 r_2} (r_1+r_2)}{c^2 M (r_1-r_2) } \frac{r_1 \Phi_{\rm t}(r_1)}
{ (r-r_1)^{3/2} (r_2-r)^{1/2}}\nonumber\\
&&\left.-\frac{\sqrt{r_1 r_2} (r_1+r_2)}{c^2 M (r_1-r_2) } \frac{r_2 \Phi_{\rm t}(r_2)}
{ (r-r_1)^{1/2} (r_2-r)^{3/2}} \right\}\nonumber\\
&=&\left( \frac{3\pi M}{c^2} -\frac{\pi b^{\rm t}_2}{ c^2 M}\right)\frac{ r_1+r_2 }{r_1 r_2} \nonumber\\
&&-\frac{3\pi b^{\rm t}_3 }{2 c^2 M }\left(\frac{ r_1+r_2 }{r_1 r_2}\right)^2.
\end{eqnarray}
Observe that this last expression is independent of $b^{\rm t}_1$.

The deflection angle, $\delta \phi$, is the shift in the orbit's perihelion and it is defined as $\Delta \phi $ minus $2\pi$, which corresponds to a closed orbit. Moreover, it is customary to write the result in terms of the length of the semimajor axis, $a$, and the orbit's eccentricity, $\epsilon$, which are such that $r_{1}=a(1-\epsilon)$ and $r_{2}=a(1+\epsilon)$. Also, units where $c=1$ are reintroduced, for simplicity. Taking all this into the account, it is possible to write
\begin{equation}
\delta \phi=\frac{ 6\pi M }{a(1-\epsilon^2)} \left(1 -\frac{ b^{\rm t}_2}{3M^2}\right) -\frac{6\pi b^{\rm t}_3 }{ M a^2(1-\epsilon^2)^2}.\label{delta phi perihelion}
\end{equation}
Relevantly, when $b^{\rm t}_2=0=b^{\rm t}_3$, the general relativistic effect is recovered. Note that $b^{\rm t}_2$ depends on $a$ and $\epsilon$ in the same way than the GR effect. On the other hand, $b^{\rm t}_3$ goes like $a^{-2}(1-\epsilon^2)^{-2}$. Thus, it is possible to set separate bounds on these nonconservation parameters using Solar system data, particularly, if such bounds were set using data of several planets. In fact, there are reported bounds for Mercury \cite{Verma} and Mars \cite{Mars}, however, such an analysis lies outside the scope of this paper.

The most stringent PPN bounds arise from studies of Mercury's orbit by the MESSENGER spacecraft \cite{Verma}. The corresponding PPN expression is \cite{Will2006}
\begin{eqnarray}
\delta \phi^{\rm PPN}&=&\frac{ 2\pi M_\odot}{a_{\mercury}(1-\epsilon^2_{\mercury})} (2 +2\gamma -\beta)\nonumber\\
&\approx& 43'' \left(1+\frac{2(\gamma-1) -(\beta-1)}{3}\right)
\end{eqnarray}
where $\gamma$ and $\beta$ are the PPN parameters and the subindex $\mercury$ stands for Mercury. In the last step, the GR value of nearly 43 arcseconds per century for the deflection angle is used, which corresponds to $2.1 \times 10^{-4}$ radians per century. In fact, these limits are actually set on $\beta$ after assuming $\gamma-1=(2.1\pm 2.3)\times 10^{-5}$ \cite{Bertotti2003}. The resulting limits are $\beta-1=(-4.1\pm 7.8)\times 10^{-5}$. With these data
\begin{eqnarray}
\delta \phi^{\rm PPN}&\approx& 43'' \left[1+(2.7\pm 4.1)\times 10^{-5}\right],
\end{eqnarray}
which, when compared to Eq.~\eqref{delta phi perihelion}, yields
\begin{equation}
 -6.8 < (3.3\times 10^4)\frac{ b^{\rm t}_2}{M_\odot^2} + 1.1 \frac{ b^{\rm t}_3 }{ M_\odot^3 }<1.4.
\end{equation}
These are the bounds on the nonconservation parameters, and they are compatible with the inequalities \eqref{condt1}-\eqref{condt4}, as required for consistency.

\section{Conclusions}\label{Conclusions}

In this paper, it is argued that matter energy conservation could be abandoned in theories that try to reconcile gravity and quantum mechanics, as in the latter theory there are processes, associated with measurements, where energy is explicitly not conserved. It is shown that, to produce a geometrical theory of gravity that is compatible with energy nonconservation, it is necessary to have either additional gravitational degrees of freedom, or nonminimal couplings, or to consider nondynamical fields.

The core of this papers is devoted to study the trajectories of pointlike particles subject to energy nonconservation. A trajectory equation is found using Papapetrou's method and the result shows that energy nonconservation generates a particular acceleration, as expected. This trajectory equation is further studied under the assumptions of staticity and spherical symmetry.

Interestingly, the unimodular theory of gravity, which is a proposal to tackle the cosmological constant problem through a nondynamical structure, satisfies two conditions that greatly simplify the trajectories' study. The first of these conditions is the existence of a generalized Birkhoff theorem that can be applied in unimodular gravity. The second condition has to do with the fact that, in unimodular gravity, the energy-momentum divergence, when viewed as a differential $1$-form, has to be closed. This, in turn, implies that the effects of energy nonconservation are encoded in a scalar function, dubbed nonconservation potential. By virtue of these results, the trajectories of light and massive particles, in the test particle approximation, can be studied using equations that resemble a system of one degree of freedom.

To probe the method, concrete nonconservation potentials are proposed for null and timelike trajectories. The deflection of light, time delay, and the shift in the perihelion are explicitly calculated, and the results are compared with Solar system data. These comparisons produce bounds on the parameters of the nonconservation potentials, implying that energy nonconservation can generate physical effects.

The ideas presented here can be generalized. Perhaps the most intriguing of such generalizations is to study the particles' trajectory equation to the next order in Papapetrou's approximation, where the (classical) spin is considered. This would allow one to test spin-dependent models of energy nonconservation, particularly those that produce an effective cosmological constant that is compatible with the observations \cite{PerezSudarsky2,PerezSudarsky}. With the methods presented here, spin-dependent models could be compared with Solar system data, producing independent tests. Note, however, that the models of Refs.~\onlinecite{PerezSudarsky2,PerezSudarsky} use the curvature scalar, $R$, as a measure for energy diffusion, and $R$ vanishes for the relevant symmetries. Still, similar models could be analyzed, where, for example, the Kretschmann scalar plays the role of $R$.

\begin{acknowledgments}
We acknowledge getting valuable feedback from H.A. Morales-T\'ecotl, U. Nucamendi, E. Okon, J.A. Garc\'ia-Zenteno, and D. Sudarsky. This research was funded by UNAM-DGAPA-PAPIIT Grant IG100120 and CONACyT FORDECYT-PRONACES Grant 140630, and also by CONACyT through the graduate school scholarships.
 \end{acknowledgments}

\bibliography{bibliography.bib}

\end{document}